\documentclass[a4paper,11pt]{article}
\usepackage{pos}
\usepackage{subcaption}
\title{Design and characterization of the Flat-Field Calibration of the NectarCAM Camera}

\author*[a,b]{Anastasiia Mikhno}
\author[a]{Federica Bradascio}
\author[a,c]{Jonathan Biteau}
\author[d]{François Brun}
\author[e]{Patrick Brun}
\author[a]{Hossam Boutalha}
\author[e]{Justine Devin}
\author[f,g]{Armelle Jardin-Blicq}
\author[h]{Pierre Jean}
\author[a]{Michael Josselin}
\author[b]{Jean-Philippe Lenain}
\author[a]{Quentin Luce}
\author[d]{Vincent Marandon}
\author[a]{Kevin Pressard}
\author[e]{Georges Vasileiadis}
\author{on behalf of the CTAO NectarCAM Collaboration}

\affiliation[a]{IJCLab, Université Paris-Saclay, CNRS/IN2P3, 91405 Orsay, France}

\affiliation[b]{LPNHE, Sorbonne Université, CNRS/IN2P3, 4 place Jussieu, 75005 Paris, France}

\affiliation[c]{Institut Universitaire de France (IUF), France}

\affiliation[d]{ IRFU, CEA, Université Paris-Saclay, 91191 Gif-sur-Yvette, France}
\affiliation[e]{LUPM, Université de Montpellier, CNRS/IN2P3, Montpellier, France}
\affiliation[f]{Université Bordeaux, CNRS, LP2I Bordeaux, UMR 5797, F-33170 Gradignan, France}
\affiliation[g]{French-Chilean Laboratory for Astronomy, IRL 3386, CNRS and Universidad de Chile, Casilla 36-D, Santiago, Chile}
\affiliation[h]{ IRAP, Université de Toulouse, CNRS, CNES, UPS, Toulouse, France}

\emailAdd{amikhno@lpnhe.in2p3.fr}

\abstract{The NectarCAM flat-field flasher is a calibration device designed for the camera that will equip the Medium-Sized Telescopes (MSTs) of the northern site of the Cherenkov Telescope Array Observatory (CTAO). Positioned in the centre of the MST dish, 16 meters in front of the camera, the flasher emits short (FWHM ${\approx}\,5$\,ns), uniform (2--4\%) light pulses to illuminate the entire focal plane.

Accurate calibration is crucial for the optimal operation of the NectarCAM, ensuring precise gain computation and mitigating differences in light-collection efficiency of the pixels of the camera. Using the flat-field flasher, two informations are obtained : the pixel gain and the relative efficiency between pixels. In addition, the flasher is used to probe the dynamic range over which the camera operates effectively. 

In this study, we report on the performance characterisation of the flat-field flasher using a dedicated test bench. We report on the results of tests conducted on several flasher units, evaluating their reliability. Furthermore, we describe how the flat-field coefficients are applied within the camera to ensure uniformity of response of few percent level across all 1855 pixels. 

As the deployment of the first MST at the CTAO northern site is scheduled for 2027, this work represents a significant contribution to the collaboration's efforts to finalize camera calibration systems. }

\ConferenceLogo{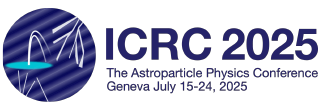}

\FullConference{39th International Cosmic Ray Conference (ICRC2025)\\
 15–24 July 2025\\
Geneva, Switzerland\\}

\begin{document}

\maketitle

\section{Introduction}

The upcoming Cherenkov Telescope Array Observatory (CTAO) will detect gamma rays at energies from a few tens of GeV up to a few hundreds of TeV.  The CTAO is going to have two sites: the northern site is located in La Palma (Canary Islands) and the southern site is located in Chile’s Atacama desert. The CTAO consists of four Large-Sized Telescopes in the northern hemisphere, up to 23 Medium-Sized Telescopes (MSTs) distributed over both array sites for its core energy range, and up to 37 Small-Sized Telescopes in the southern hemisphere.
The CTAO will significantly improve our understanding of the gamma-ray sky above 30\,GeV, with an effective area several times larger than that of the current generation of imaging atmospheric Cherenkov telescopes (IACTs), depending on energy.

{The NectarCAM \cite{NectarCAM_intro, 2020NIMPA.95062949B} is the camera that will equip all the 9 MSTs of the northern site of the CTAO. The NectarCAM features 1855 pixels using high-sensitivity  photomultiplier tubes (PMTs) wavelength range ~300–600\,nm, and provides two gain channels  for recording fast Cherenkov signals with 1\,GHz sampling over 60\,ns windows. The high gain channel is meant for measurements at the few photoelectron level, whereas low gain channel is used for reconstruction up to 2000 photoelectrons (PE).

Accurate calibration is essential for the optimal performance of the NectarCAM. To interpret the raw signals recorded by each pixel of the camera into physical quantities related to observed gamma-ray, the camera must undergo a series of calibration procedures. These ensure consistent response across all pixels, account for the effects of electronics and optics, and enable precise reconstruction of the incident Cherenkov light. This section outlines part of the calibration strategy and data processing pipeline developed to validate the performances of the NectarCAM, with a focus on pedestal, gain, and flat-field coefficient (FF) calibration. Currently, the NectarCAM is being tested in the dark room at CEA Irfu to verify all the calibration systems and to show compliance with CTAO requirements.

The calibration of the NectarCAM involves several distinct steps:

\begin{itemize}
    \item \textbf{Pedestal Calibration:} Measures the fluctuations of the baseline of the signal (in analog-to-digital converter -- ADC -- units). The pedestal includes contributions from non-zero offset with electronic noise and night sky background during observations. Measurements under dark conditions, with the camera shutter closed, allows us to isolate this electronic noise component. For the NectarCAM, the typical mean of the pedestal baseline is approximately 250 ADC, though this value may vary due to thermal effects during electronics warm-up, night sky background and ambient temperature changes. 

    \item \textbf{Gain Calibration:} Converts the ADC signal into units of PE for each readout channel. This is crucial for the extraction of the relevant signal, or charge, of each pixel of the camera, and thus crucial  for the energy reconstruction of the gamma-ray.
    \item \textbf{Flat-Field Coefficient Calibration:} Even for a uniform incident light front, a pixel-dependent response is expected due to non-uniformities in Winston cones and photodetection efficiency. The flat-field calibration corrects for these discrepancies in signal amplitude, ensuring a homogeneous camera response.

For the FF calibration, the flasher is positioned in the centre of the MST dish, about 16 meters in front of the camera. The flasher emits short (FWHM ${\approx}\,5$\ ns), uniform (2--4\%) light pulses, at a wavelength of 390\,nm, to illuminate the entire focal plane.
\end{itemize}

A run is a set of triggered events acquired under fixed, well-defined settings. In this study, each run consisted of 10,000 events recorded at CEA Irfu. For the FF flasher calibration, three different LED configurations were tested, varying in intensity and illumination pattern (a detailed description is provided in Section~\ref{sec:LEDs}).

The  light pulses emitted by the flasher are observed by each pixel. It  results in a waveform trace, i.e. time series of voltage samples (in ADC  counts), which encodes the raw signal information prior to pedestal subtraction, gain correction and charge extraction. This is defined as the R0-level data in the the  \textbf{NectarChain} pipeline \cite{Grolleron:2023T8}. The pipeline is then used to transform the R0-data into calibrated, analysis-ready data,  R1-level data.

The charge extraction from the waveforms is performed using the \textbf{ImageExtractor} class of ctapipe \cite{ctapipe-icrc-2023}. It applies a charge integration algorithm  looking for the peak of the maximum signal and integrating it on a window of $-4\pm8$ ns. The integrated charge in PE which accounts for the gain. In this work we used the Global Peak Window Sum integration algorithm, with corresponding window width of 5\,ns.

As an initial step in the flat-field calibration, it is essential to characterize the performances and stability of the flasher. 
In the runs used in this study, the flasher sent periodic pulses at 1\,kHz frequency. We analysed the time intervals between consecutive flashes and confirmed their stability over the duration of test runs. The mean inter-flash time was measured to be 
$\langle \Delta t \rangle = 1000.1 \pm 7.5~\mathrm{\mu s}$.

\begin{figure}[htbp]
    \centering
    \includegraphics[width=0.5\textwidth]{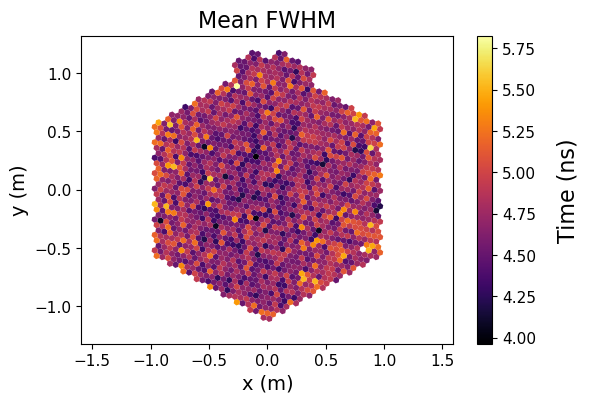} 
    \caption{Mean FWHM of pulses from the FF flasher as a function of the pixel
coordinates in the camera frame. Note that most of the outer ring modules were not installed during these measurements, except for five modules on the upper part of the ring.}
    \label{fig:FWHM}
\end{figure}

 Additionally, we examined the mean waveform response of  pixels. From the mean waveform computed for $10^4$ events, we measure the mean rise time defined as $t_{10\%}-t_{90\%}$, where $t_{10\%},  t_{90\%}$ are  the time at which $10\%$, $90\%$ of the maximum amplitude is reached before the maximum of the trace, and the mean decay time defined as $t_{90\%}-t_{10\%}$ after the maximum of the trace. The FWHM distribution over the pixels of the partially equipped camera is shown in Fig~\ref{fig:FWHM}. The white pixels in this plot, as well as in the others, indicate pixels with invalid values (e.g., gain equal to zero). The shape of the camera image results from the absence of focal plane modules (PMTs) in the outer row.

\section{Computation of FF coefficients}

To accurately compute the illumination of the NectarCAM, a reliable estimation of the gain is required. This is crucial for interpreting the recorded signal in terms of physical light intensity, and for enabling the following steps in the calibration pipeline, such as FF. This study is aiming to improve the FF calibration by taking into account the flasher’s light-front shape. The requirements are to control the light front with a precision better than 2\%.

\subsection{Photostatistics Method}

For this purpose, the photostatistics method has been employed as a well-established technique in Cherenkov camera calibration \cite{photostat_beta}. The key idea behind this algorithm is to estimate the absolute gain by analysing the statistical moments of the recorded signal distribution across pixels. Specifically, the method relies on the assumption that the response (in PE) to light pulses follows a Poisson process. Under this assumption, the first and second-order moments of the recorded charge distribution (after pedestal subtraction) carry the needed information to infer the gain.

The absolute gain $g$ is expressed as:

\begin{equation}
    g = \frac{\sigma_x^2 - \sigma_{\text{ped}}^2}{\langle x \rangle (1 + \beta^2)},
    \label{photostat}
\end{equation}

where  $\sigma_{\text{ped}}^2$ is the variance of the pedestal distribution (in ADC),
$\sigma_x^2$ is the measured variance of the signal distribution (in ADC),  $\langle x \rangle$ is the mean recorded signal (in ADC) and $\beta$ represents the width of the single photon response (Single PE, SPE) distribution, expressed relative to the gain. 
The factor $\beta$ accounts for the fluctuations in the SPE response due to intrinsic variations in the photon detection process (including electron multiplication in PMTs). It is determined via a fit to the SPE distribution using dedicated single-photon runs. Furthermore, in this formula, the term corresponding to the standard variation due to flasher variability is neglected. This holds as the variance from Poisson photoelectron statistics typically dominates the signal, making the smaller contribution from flasher variability negligible in comparison. 

The value of the coefficient $\beta$ is expected to remain stable for a given PMT \cite{photostat_beta} . Consequently, it is often sufficient to compute $\beta$ once for a given set of calibration runs taken under similar conditions (e.g., same flasher intensity and environmental settings), rather than recomputing it for each individual event or pixel.

\subsection{Characterization of the flasher's light-front}

To model the illumination pattern produced by the flat-field flasher for each pixel of the camera, we represent the expected light distribution as a two-dimensional Gaussian function. This assumption is motivated by the quasi-symmetric and smooth nature of the light front emitted by the flashers.

The Gaussian model is parametrized by its amplitude $A$, centre position $(x_0, y_0)$, and width parameter $\sigma_{xy}$  along both the x and y directions:

\begin{equation}
    f(x, y) = A \cdot \exp\left( -\left( \frac{(x - x_0)^2 + (y - y_0)^2}{2\sigma_{xy}^2} \right) \right),
    \label{Gaussian}
\end{equation}

The function $f(x_i, y_i)$ is evaluated at the position of each pixel $(x_i, y_i)$ to obtain the expected photoelectron count per pixel. This model serves as the baseline prediction to which we compare the measured data in a fit procedure.

To determine the best-fit parameters $(A, x_0, y_0, \sigma_{x,y})$, we perform a chi-square minimization using the \texttt{Minuit}\footnote{\url{https://iminuit.readthedocs.io/}} optimization package. The chi-square function is defined as:

\begin{equation}
    \chi^2 = \sum_i \frac{(s_i - f_i)^2}{\sigma_i^2},
\end{equation}

where $s_i$ is the measured signal in the $i$-th pixel (in PE),  $f_i = f(x_i, y_i)$ is the model prediction for that pixel,  $\sigma_i$ is the uncertainty on the PE count in the pixel.

\begin{figure}[htbp]
    \centering
    \includegraphics[width=1\textwidth]{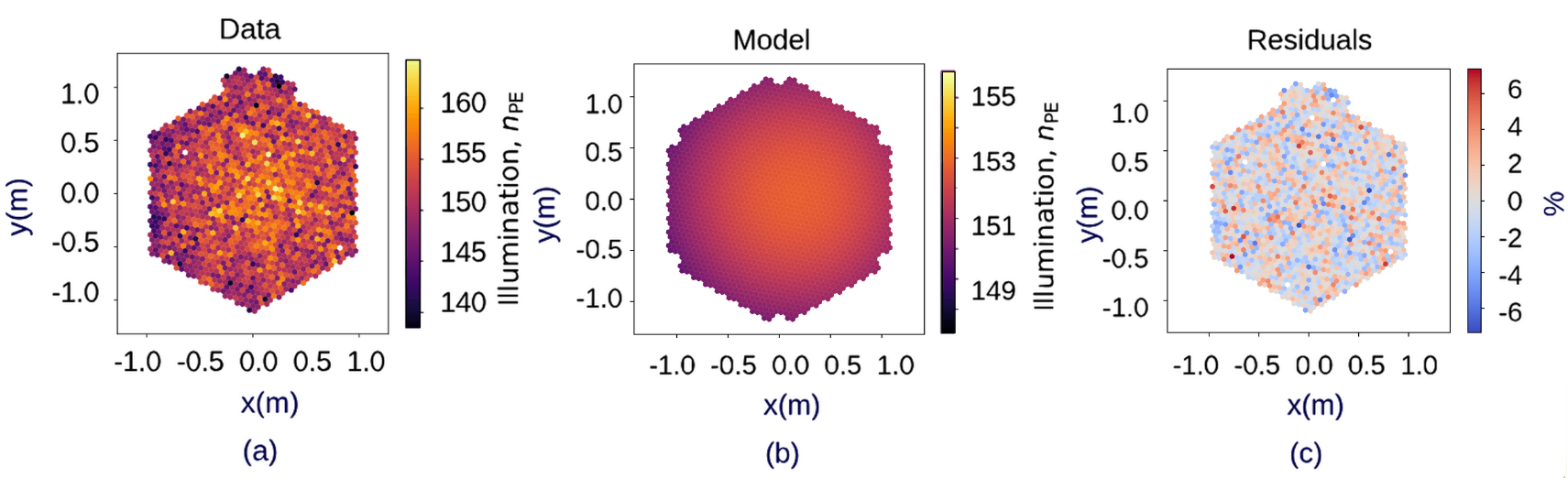} 
    \caption{Data from the run, 2D Gaussian best fit, and corresponding residuals. In the panel \textbf{(a)} the illumination of the camera is plotted as a function of pixel coordinates, \textbf{(b)} the fitted illumination (in PE) is presented as a function of pixels coordinates, in panel \textbf{(c)} the corresponding residuals between data and the model is presented as a function of pixels coordinates. Data were taken at 9V, with the main LED configuration.}
    \label{fig:waveform}
\end{figure}

Prior to the analysis, we verified on independent flasher runs that splitting a long run of 90,000 recorded events into ten smaller segments of 9,000 events each yields consistent results, confirming that datasets of this size are sufficient to ensure reliable model fitting. 

The analysis was performed for 3 different configurations of the flasher's LEDs, under different voltages from 9V to 15V. In total, this resulted in 21 datasets for analysis.  We exemplify the discussion here on the data taken at 9V.

\begin{figure}[htbp]
    \centering
    \includegraphics[width=0.5\textwidth]{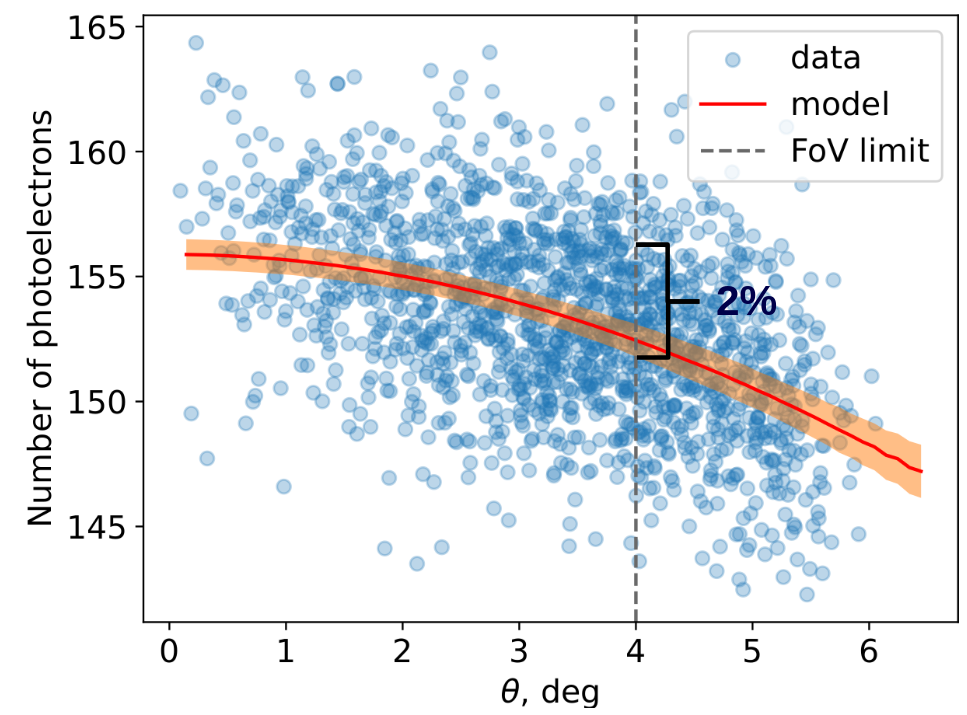} 
    \caption{Radial distribution of detected intensity. Intensity is given as a function of angular distance from the peak of the fitted Gaussian. Data points are in blue, the fitted model is in red with fit uncertainties given in orange band, the gray dashed line shows the end of the field of view of NectarCAM 16m away from the flasher. The data taken in the Irfu dark room were obtained for a distance of 12m. Data were taken at 9V, main LED configuration.}
    \label{fig:model_run_6241}
\end{figure}

In the Fig.~\ref{fig:waveform}, the result of the fit of a 2D Gaussian model to the data from a run taken at 9 volts is presented along with the residuals, which are centred around zero and go up to $6\%$ . In the Fig.~\ref{fig:model_run_6241}, the same fit result is presented as a radial profile, plotting the pixel values as a function of their angular distance from the centre of the fitted Gaussian. Here, the red line corresponds to the fit, and the blue points represent the data. We obtained an uncertainty on the fitted model of 0.4\%.  It was possible to probe angular distances $\theta > 4$ deg because the distance at Irfu is only 12m, whereas for the MSTs it is going to be 16m.

\subsection{Flat-Field Coefficients}

To correct for inhomogeneities in pixel-to-pixel response across the camera, we compute FF coefficients. These coefficients normalize the pixel amplitudes to a reference level and are essential for achieving a uniform photometric response across the focal plane.

The flat-field coefficient for a given gain channel $i$ and pixel $j$ is defined as:

\begin{equation}
    C_\text{FF}(i,j) = \frac{A_{\text{pix}}(i,j)}{\mu_{\text{cam}}(i)},
    \label{eq:FF_coefs}
\end{equation}

where  $A_{\text{pix}}(i,j)$ is the charge for pixel $j$ in gain channel $i$ (in PE),  $\mu_{\text{cam}}(i)$ is the expected amplitude in channel $i$. In Fig.~\ref{fig:model_run_6241}, $\mu_{\text{cam}}(i)$ corresponds to the best fit values (red line).

The ratio in Eq.~\ref{eq:FF_coefs}  effectively rescales each pixel to the average response level of its gain channel. Accounting for the Gaussian illumination pattern ensures that the flat-field coefficients reflect the true camera response.

\begin{figure}[htbp]
    \centering
    \begin{subfigure}[b]{0.4\textwidth}
        \centering
        \includegraphics[width=\textwidth]{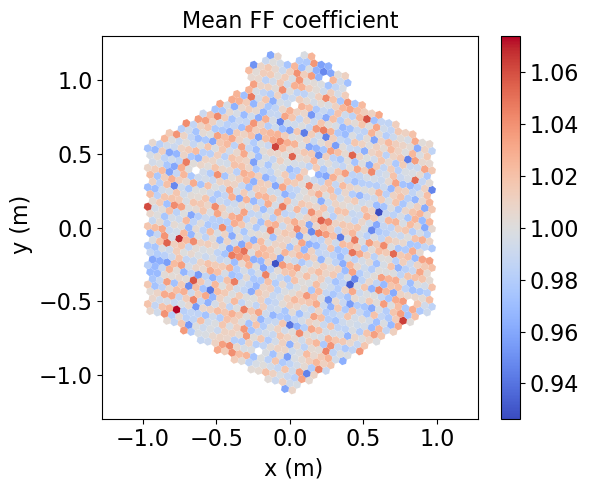}
        \caption{Flat-Field coefficients over pixel coordinates.}
        \label{fig:FF_upd}
    \end{subfigure}
    \hfill
    \begin{subfigure}[b]{0.55\textwidth}
        \centering
        \includegraphics[width=\textwidth]{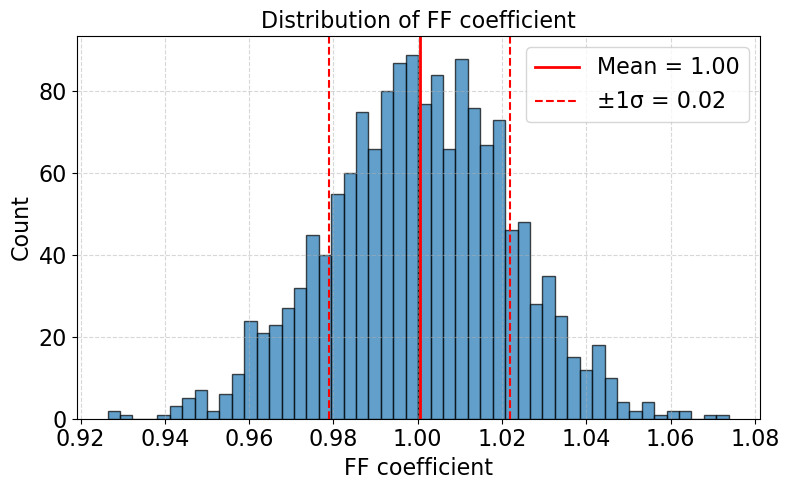}
        \caption{Distribution of Flat-Field coefficients.}
        \label{fig:ff_hist}
    \end{subfigure}
    \caption{Flat-Field coefficients computed for the run taken at 9V, main LED configuration. (a) Coefficients as a function of pixel coordinates. (b) Histogram of their distribution over the camera.}
    \label{fig:ff_combined}
    
\end{figure}

For the main LED configuration of the flasher, FF coefficients were computed at different voltage regimes (9-15 V). In Fig.~\ref{fig:ff_combined} the result is shown for the run taken at 9 V. The standard deviation of the FF coefficients for the run is $0.02$.

\subsection{LED configurations and IJCLab test bench}
\label{sec:LEDs}
The runs analysed  in these proceedings have been taken in the integration hall at CEA Irfu. In parallel, at the ĲCLab test bench, all the flashers are tested using one camera module to probe the flasher's light front while cancelling the effects which might appear due to differences in camera pixels.
Each flasher is being tested at different voltage regimes, and different LED configurations are probed (see Fig.~\ref{fig:LEDs}). Three configurations were selected for the flat-field flasher to ensure operational reliability in case of LED failures. The main configuration uses a set of LEDs chosen to produce the most homogeneous illumination across the camera. The backup configuration provides an alternative LED set that can be used if some LEDs in the main setup fail, while maintaining a well-characterized light-front. The low-illumination configuration delivers a reduced light intensity for calibration at lower signal levels.

\begin{figure}[htbp]
    \centering
    \includegraphics[width=0.5\textwidth]{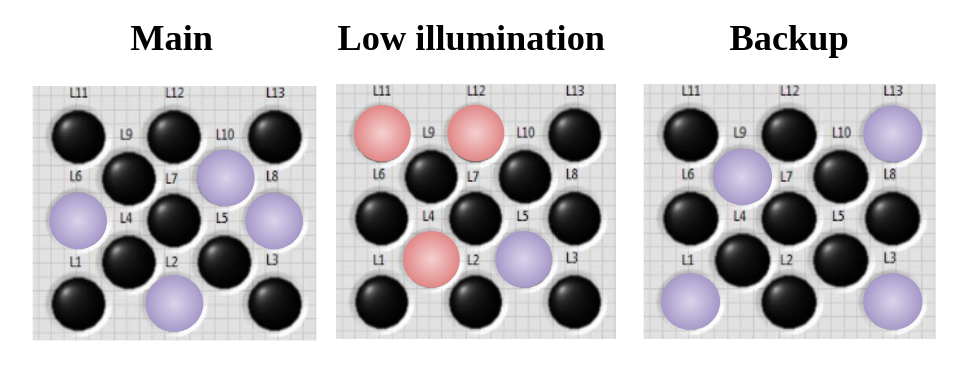} 
    \caption{Configurations of the LEDs of the flasher. Black circles correspond to the LEDs which are turned off in this configuration, purple circles represent LEDs which are turned on, the red circles are LEDs which produce a lower illumination when turned on.  }
    \label{fig:LEDs}
\end{figure}

In Fig.~\ref{fig:Quentins_tension_vs_charge}, the results of the voltage tests are illustrated for different configurations of the flasher's LEDs. For all configurations, the rise in illumination was observed with increasing voltage of the the LEDs, and the PMTs were saturated at higher voltages analogously to the results obtained at the CEA test bench.  
Furthermore, the light front was fitted and compared for both test benches. The resulting Gaussian widths are the following: $\sigma = 0.1928 \pm 0.0003 ~(\rm rad)$ for the IJCLab test bench, and  $\sigma = 0.19 \pm 0.01 ~ (\rm rad)$ for data obtained at CEA, which proves that we can characterize the light front using the data from camera runs.
 
\begin{figure}[htbp]
    \centering
    \includegraphics[width=0.5\textwidth]{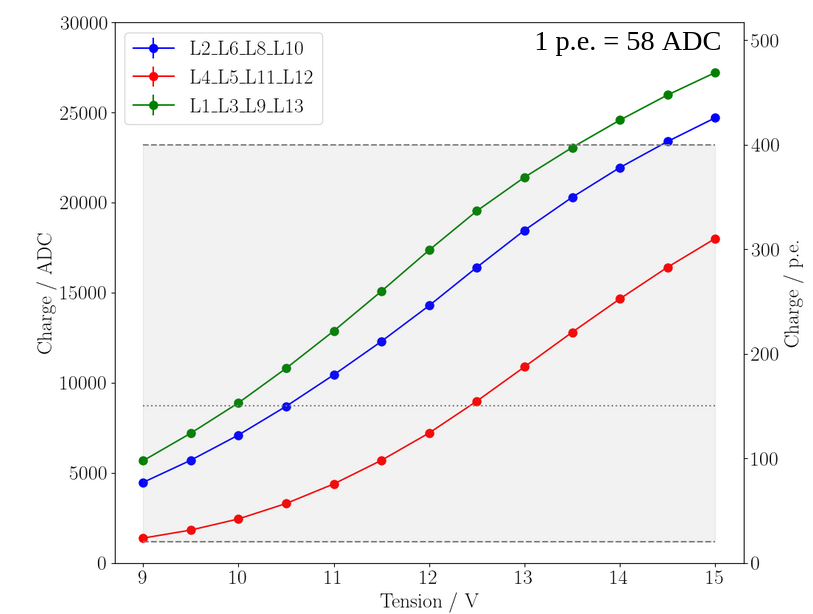} 
    \caption{Dependence of charge on the voltage for different LED configurations of the flasher: green line corresponds to the backup LED configuration, blue line corresponds to the main LED configuration, and the red line to the low illumination configuration. }
    \label{fig:Quentins_tension_vs_charge}
\end{figure}

\section{Conclusion}

We presented the design and validation of the FF calibration system for the NectarCAM camera prototype of the CTAO MSTs. Test bench measurements confirmed the flasher's stability, delivering short ($\simeq$5\,ns), spatially uniform (2--4\%) light pulses across the focal plane. 

We developed and validated the FF calibration system for NectarCAM. A 2D Gaussian was fitted to the light distribution to extract FF coefficients, with a few-percent precision across the camera. The photostatistics method was applied using data from both the primary test bench and a parallel setup in Paris-Saclay, showing good agreement between measurements. These results confirm the reliability of calibration system for integration into the NectarCAM pipeline.

\section{Acknowledgments}

We gratefully acknowledge financial support from the agencies and organizations listed here:   
https://www.ctao.org/for-scientists/library/acknowledgments/

\end{document}